\documentclass[showpacs,preprintnumbers,amsmath,amssymb, prd,11pt]{revtex4}
\usepackage{amsmath}
\usepackage{graphicx}
\usepackage{amssymb}
\usepackage{amsfonts}
\usepackage{latexsym}
\def\beq{\begin{eqnarray}}
\def\eeq{\end{eqnarray}}
\def\ln{\,\mbox{ln}\,}

\def\al{\alpha}
\def\be{\beta}

\def\de{\delta}
\def\vp{\varepsilon}
\def\ep{\epsilon}

\def\la{\lambda}
\def\na{\nabla}
\def\pa{\partial}

\def\si{\sigma}

\def\ph{\varphi}
\def\ta{\tau}

\def\Ga{\Gamma}

\def\La{\Lambda}

\def\Om{\Omega}

\begin{document}
\vskip 0.7cm
\title{Vacuum stress-tensor in SSB theories}
%%%%%%%%%%%%%%%%%%%%%%%%%%%%%%%%%%%%%%%%%%%%%%%%%%%%%%%%%%%%%%%%%%
\author{Manuel Asorey}
\email{asorey@saturno.unizar.es}
\affiliation{Departamento de Fisica Teorica, Universidad de Zaragoza,
    50009, Zaragoza, Spain}
\author{Peter M. Lavrov}
\email{lavrov@tspu.edu.ru}
\affiliation{Tomsk State Pedagogical University. Kievskaya 60, 634061,
Tomsk, Russia}
\author{Baltazar J. Ribeiro}
\email{baltazarjonas@fisica.ufjf.br}
\author{Ilya L. Shapiro}
\email{shapiro@fisica.ufjf.br}
\altaffiliation{Also at Tomsk State Pedagogical University, Tomsk, Russia.}
\affiliation{Departamento de F\'{\i}sica, ICE,
Universidade Federal de Juiz de Fora, 36036-330, MG, Brazil}

%%%%%%%%%%%%%%%%%%%%%%%%%%%%%%%%%%%%%%%%%%%%%%%%%%%%%%%%%%%%%%%%%%
\begin{abstract}
The renormalized energy-momentum tensor of vacuum has been deeply
explored many years ago. The main result of these studies was that
such a tensor should satisfy the conservation laws which reflects
the covariance of the theory in the presence of loop corrections.
In view of this general result we address two important questions,
namely how to implement the momentum cut-off in a covariant way and
whether this general result holds in the theory with Spontaneous
Symmetry Breaking. In the last case some new interesting details
arise and although the calculations are more involved  we show that
the final result satisfies the conservation laws.
\end{abstract}
%%%%%%%%%%%%%%%%%%%%%%%%%%%%%%%%%%%%%%%%%%%%%%%%%%%%%%%%%%%%%%%%%
%%%%%%%%%%%%%%%%%%%%%%%%%%%%%%%%%%%%%%%%%%%%%%%%%%%%%%%%%%%%%%%%%%
\maketitle
\section{Introduction}

Traditionally, the calculation of quantum corrections to the
stress-tensor (also called Energy and Momentum Tensor - denoted
as EMT in what follows) of vacuum is one of the most important
issues of Quantum Field Theory in curved space-time. The reasons
for the special interest is this problem are becoming obvious
if we remember that the matter fields and particles enter the
cosmological and most other gravitational equations in the form
of EMT of matter, which is usually taken as a fluid. The quantum
effects of field fluctuations turn out to give some corrections
to the corresponding equations of state. The most relevant
example, probably, is that the EMT for radiation gains a non-zero
trace due to the conformal (trace) anomaly, which really changes
the equation of state for the radiation with some possible
relevant effects for the radiation-dominated Universe
\cite{RadiAna}. Opposite to this case, the equation of state for
the massive particles and baryonic matter in general, does not
change essentially, because quantum corrections can not make
such a matter content to be relativistic.

The situation is quite different in the case of vacuum quantum
effects, which can be much more relevant than those of the
matter sector. Recently there were many publications on this
subject, including the ones where the possible quantum effects
of quantum massive matter fields on cosmology and astrophysics
were explored. In particular, it was noticed in
\cite{CC-nova,babic,CCfit,Gruni,Stefan,LRL} that such quantum
corrections can be defined up to a single free parameter
$\nu$ on the background of general covariance. Some
observational consequences of the possible quantum corrections
were explored in \cite{CCwave,CCG} and led to establishing an
upper bound on the magnitude of $\nu$. Furthermore, the same
unique form of quantum corrections was applied also to
astrophysics \cite{Gruni} and was shown \cite{RotCurves} to
provide an accurate description of the rotation curves for
some sample set of disk galaxies, without introducing a
large amount of Dark Matter content (see \cite{QFEXT-11-proc}
for more examples). Some other applications to cosmology and
astrophysics were also discussed in Ref. \cite{AlFa}.

The applications mentioned above are based on a single,
however nontrivial assumption of the existence of relevant
quantum corrections in the low-energy vacuum sector. Needless
to say that the most desirable development would be to derive
such quantum effects on the regular basis in the framework of
some rigorous QFT approach. The problem was discussed in
\cite{DCCR} and the final conclusion concerning existing regular
methods was essentially negative. The required quantum correction
to the effective action of vacuum should be given by a sum of
infinite products of the curvature tensor components with an
infinite number of non-local insertions, hence there are small
chances for a practical realization of such a calculus. One can
note that the situation becomes much more definite if we give up
the covariance and use, for instance, the conformal parametrization
of the background metric. In this case it is possible
to calculate the  quantum corrections \cite{Shocom}. However, this
method is not really safe and is anyway applicable only at
the high-energy regime when the minimal subtraction procedure
is supposed to be reliable.

It would be very nice to have some alternative approach to the
derivation of desirable quantum corrections. Recently there were
some publications where the result was obtained by means of the
cut-off regularization in the conformally flat cosmological
metric case \cite{Magg1} (early version) and \cite{Bilic} (see
also \cite{Prokopec}).
The main idea is to perform calculations of the ``energy
density'' and ``pressure'' of the vacuum in the momentum
cut-off regularization, taking the expansion of the Universe
into account perturbatively, order by order in the Hubble
parameter $H$. The zero-order approximation has been considered
before by Akhmedov in \cite{Akhmedov} and earlier by DeWitt
\cite{DeWitt-75}. The output of the non-covariant procedure
is not the naively expected equation ``equation of state''
\ $p_{vac} = - \rho_{vac}$ \ of the cosmological constant, but
the one for the radiation  $p_{vac} = \rho_{vac}/3$, which
led to several attempts to understand this result and even
to correct it at the {\it ad hoc} basis \cite{Sirlin}. In fact,
DeWitt explained the result in a very general terms as being
produced by the non-covariant regularization. The calculations
of \cite{Magg1,Bilic} were based on the subtraction of the
flat-space result of \cite{Akhmedov}, which led to the new
``equation of state'' for the vacuum, this time proportional
to $H^2$ times the square of the cut-off parameter. The main
problem with this result is that it apparently contradicts
either the general covariance of the effective action, or
the locality of the requested counterterms. in this case we
meet a violation of the well established fundamental features
of renormalization in curved-spaces (see, e.g., books
\cite{birdav,book} and recent papers \cite{Ren-Curved}).
However, the results of these calculations should be
considered as a motivation for the study of possible existence
of the \ ${\cal O}(H^2)$-type corrections to the vacuum energy
in cosmology. At the same time it looks very important to
better understand these results at the technical level. This
consideration is one of the motivations for the present paper.
Furthermore, it is interesting to see how the calculations in
the cut-off regularization can be done covariant. This problem
has been recently solved in \cite{BaDoGu} on the basis of local
momentum representation in Riemann normal
coordinates (alternatively, one can achieve the
covariance of finite expressions by imposing the conservation
law step by step when adding specially adjusted non-covariant
counterterms \cite{Magg1,Magg2}). Furthermore, there is
one more possibility which deserves to be checked in full
details. The cosmological constant term consists of the
two main contributions
\cite{weinberg89}, namely the vacuum classical term and
the induced term. The no-go statement of \cite{DCCR} concerns
only the quantum contribution to the vacuum part and, therefore,
there is a chance to meet ${\cal O}(H^2)$-type quantum corrections
to the vacuum energy from induced part. As one can see in what
follows, for the induced contribution the route from effective
action to the EMT is not so direct as it is for the vacuum
counterpart. The corresponding calculation requires more efforts
and concerns the main purpose of the present paper. We shall
derive the quantum contribution to EMT from the induced term
in the covariant way, in the linear in curvature approximation
and will eventually show that in this approximation EMT of vacuum
is local, satisfies the conservation law and hence it is given by
a linear combination of the metric and Einstein tensor.

The paper is organized as follows. In Sect. 2 we present a brief
summary of renormalization in curved space-time and discuss the
non-covariant results obtained on the cosmological background from
this perspective. In Sect. 3 we consider, following \cite{Sponta},
the spontaneous symmetry breaking in curved space and derive the
corresponding classical vacuum EMT in the linear in curvature
approximation. Sect. 4 is devoted to the conservation law for
the EMT of the vacuum in the theories with SSB. In Sect. 5 we
present as additional technical discussion of the classical
EMT of the vacuum and its physical relevance in different theories.
In Sect. 6 we derive the one-loop quantum correction to this EMT.
Finally, in Sect. 7 we draw our conclusions and present some
additional discussions.  Some calculations concerning the normal
coordinates and local momentum representation are addressed in
Appendix A and a detailed derivation of equations of motion in
the linear in curvature approximation is contained in Appendix B.

%%%%%%%%%%%%%%%%%%%%%%%%%%%%%%%%%%%%%%%%%%%%%%%%%%%%%%%%%%%%%%%%
%%%%%%%%%%%%%%%%%%%%%%%%%%%%%%%%%%%%%%%%%%%%%%%%%%%%%%%%%%%%%%%%
%%%%%%%%%%%%%%%%%%%%%%%%%%%%%%%%%%%%%%%%%%%%%%%%%%%%%%%%%%%%%%%%
%%%%%%%%%%%%%%%%%%%%%%%%%%%%%%%%%%%%%%%%%%%%%%%%%%%%%%%%%%%%%%
\section{Brief summary of renormalization in curved space}
%%%%%%%%%%%%%%%%%%%%%%%%%%%%%%%%%%%%%%%%%%%%%%%%%%%%%%%%%%%%%%

The renormalization of quantum theory of matter fields in curved
space-time was subject of many investigations starting from
\cite{UtDW}. The most simple way to remove divergences by the
consistent renormalization procedure is related to the effective
action method \cite{buch84,Toms,book} (including by means of
Batalin-Vilkovisky formalism \cite{Ren-Curved}). The result
of all these studies can be formulated in a simple form as
follows: the theory of quantum
matter fields which is renormalizable in flat space can be
formulated as renormalizable in curved space if there is a
regularization which is consistent with general covariance from
one side and the gauge symmetries of the theory from another one.
The renormalizability means that the divergences of effective
action (at any loop order)  are local and
general covariant expressions compatible with the given gauge
symmetries.

From the effective action perspective the renormalization of  EMT
is looking quite trivial: one has to derive the effective action
\ $\Ga$ \ and take the variational derivative
\beq
\langle T_{\mu\nu}(x) \rangle \,=\,
-\,\frac{2}{\sqrt{-g(x)}}\,g_{\mu\al}(x)\,g_{\nu\be}(x)
\,\frac{\de \Ga}{\de g_{\al\be}(x)}\,.
\label{Tmn}
\eeq
After that one has to introduce the counterterms into the effective
action and add them to the $\Ga$ in (\ref{Tmn}), which equivalent
to performing some very special subtraction of the divergent terms.
This subtraction should exactly correspond to the covariant and
local counterterms in the effective action. After that the
coefficients of the remaining finite terms should be fixed by
imposing the renormalization conditions on the renormalized
classical action and/or renormalized EMT. For this end such a
classical action should be chosen in a special way and include
all the structures which are possible to emerge as counterterms.

The arguments based on covariance, locality and power counting
lead to the following form of the classical action of external
metric (vacuum):
\beq
S_{vac} &=& S_{EH}\,+\,S_{HD}\,,
\label{vacuum}
\eeq
where $\,S_{EH}\,$ is the Einstein-Hilbert action with
the cosmological constant
\beq
S_{EH}
\,=\,-\,\frac{1}{16\pi G}\int d^4 x\sqrt{-g}\,
\left(\,R+2\La\,\right)\,.
\label{EH}
\eeq
and
\beq
S_{HD} &=& \int d^4x \sqrt{-g}
\left\{a_1C^2+a_2E+a_3{\Box}R+a_4R^2 \right\}\,.
\label{HD}
\eeq
Here $\,C^2=R_{\mu\nu\al\be}^2 - 2 R_{\al\be}^2 + (1/3)\,R^2\,$
is the square of the Weyl tensor and
$\,E = R_{\mu\nu\al\be}^2 - 4 R_{\al\be}^2 + R^2\,$
is the integrand of the Gauss-Bonnet topological term (Euler
density in $d=4$). Let us remark that the presence of higher
derivative terms and cosmological constant are necessary to
have a renormalizable theory.

In the present paper we will be interested to perform covariant
calculations around the flat space-time in the linear in curvature
approximation. This means we will systematically ignore the
higher derivative part (\ref{HD}) and, in general, will not pay
attention to the ${\cal O}(R_{...}^2)$ and ${\cal O}(\Box R)$-terms.
This means, in particular, that the form of the divergent structures
which one can meet in $\langle T_{\mu\nu} \rangle$ is restricted
to the two terms, namely the ones proportional to $g_{\mu\nu}$
which are responsible for the renormalization of the cosmological
constant term and the ones proportional to the Einstein tensor
$G_{\mu\nu}=R_{\mu\nu}-(1/2)Rg_{\mu\nu}$ and responsible for the
renormalization of the Einstein-Hilbert term in the effective action.

Our calculations will be performed in the local momentum
representation, based on the use of Riemann normal coordinates.
Also, we shall use very simple cut-off regularization in the
Euclidean local momentum space. This regularization has been
shown equivalent to the cut-off of the proper time integral in
the Schwinger formalism in flat space \cite{Liao} and recently
has been used in \cite{CorPot} to calculate effective potential
of the scalar field in curved space-time.

An alternative approach to renormalize EMT in curved space-time
is to work directly with the classical expression for the EMT and
perform calculation. This approach is the most traditional one (see
\cite{birdav} and references therein). The covariant calculations
in this way have been performed in \cite{Chris76} and \cite{Chris78}
by means of the point-splitting regularization, without or with the
use of effective action method. The covariant structure of divergences
of EMT which has been described before is restored in the limit
of zero splitting, but only if this limit is taken in a special
invariant way.

Let us consider the result of \cite{Chris76} for the quartic
divergent  part of the quantum corrections
to EMT. For the sake of simplicity we can deal with the flat space
expressions, because the quartic divergences are not really affected
by this choice. Then
\beq
\langle T_{\mu\nu}\rangle_{\mbox {quart. div}}
&=&
\frac{1}{2\pi^2}\, \frac{1}{n_\al n^\al}\,
\Big( g_{\mu\nu} - 4 \,\frac{n_\mu n_\nu}{n_\be n^\be}\Big)\,,
\label{a1}
\eeq
where $n_\al$ is a small non-null four-vector defining the point
splitting regularization of the corresponding Green functions
\ $G(x,x)\to G(x, x+n)$. Now, if we chose the vector \ $n$ \ in
temporal direction \ $n=(\epsilon^2,0.0.0)$ \ we get that
\beq
\langle T_{\mu\nu}\rangle_{\mbox {quart. div}}
&=& -\,\frac{1}{2\pi^2\ep^4}
\left(
\begin{array}{cccc}
3  &  0  &  0  & 0  \\
0  &  1  &  0  & 0  \\
0  &  0  &  1  & 0  \\
0  &  0  &  0  & 1  \\
\end{array}
\right)
\label{a2}
\eeq
which is a traceless quartic divergent component of the total
energy-momentum tensor. This result is in accordance with the
one based on naive momentum cut-off in flat space of \cite{Akhmedov}.
As it was explained in \cite{DeWitt-75}, there is no contradiction
with the expected local Lorentz invariance of the divergences,
because the origin of the (\ref{a2}) is in the use of a
non-covariant regularization. In case of the point-splitting
with temporal direction the breaking of Lorentz invariance is
due to the non-relativistic choice  \ $n=(\epsilon^2,0.0.0)$.
In case of cut-off regularization the origin of a non-covariance
is different but since it is equally non-relativistic, the
final result is the same.

One  expects that a  Lorentz invariant regularizations  would give
rise to \cite{Chris76} \cite{Chris78}
\beq
\langle T_{\mu\nu}\rangle_{\mbox {quart. div}}
&=& \frac{1}{2\pi^2\ep^4}
\left(
\begin{array}{cccc}
1  &   0  &  0  &  0   \\
0  & - 1  &  0  &  0   \\
0  &   0  & -1  &  0   \\
0  &   0  &  0  & -1   \\
\end{array}
\right)\,,
\label{a3}
\eeq
which is proportional to the Minkowski metric tensor and can be
interpreted as a standard divergent contribution to the zero
point energy or cosmological constant term.

Let us show how this occurs in   Pauli-Villars regularization
\cite{ags}. The contribution of a free scalar field with mass
$m$ to the vacuum energy $\rho$ is given by \cite{Akhmedov}
\beq
\rho=\frac1{16\pi^2}\left(  \Omega^4+ m^2 \Omega^2
+ \frac1{8} m^4 - \frac1{2} m^4
\log\frac{2\Omega}{m}+{\cal O}\left(\frac{m}{\Omega}
\right)\right),
 \eeq
 where $\Omega$ is a  3-momentum space cut-off.
 Whereas, the same contribution to the pressure reads,
  \beq
 p=\frac1{48\pi^2}\left(  \Omega^4- m^2 \Omega^2
 - \frac{7}{8} m^4 + \frac{3}{2} m^4 \log\frac{2\Omega}{m}
 +{\cal O}\left(\frac{m}{\Omega}\right)\right).
 \eeq

  The Pauli-Villars regularization is defined by a family of scalar
  and ghost fields with masses
  $m_i^2= \mu_i^2 M^2 + m^2, i=1, 2,\cdots N$
  with degeneracies $s_i$. Positive degeneracies correspond to scalar
  fields and negative degeneracies to ghost fields.
  The Pauli-Villars conditions
    \beq
    \sum_{i=1}^N s_i= -1, \qquad
    \sum_{i=1}^N s_i \mu_i^2= 0,
\qquad
    \sum_{i=1}^N s_i \mu_i^4= 0
    \label{pvc}
      \eeq
 guarantee that for a free field theory with mass $m$ the
 quantum corrections to the vacuum energy and pressure
 are finite, i.e. all $\Om$ quartic, quadratic and logarithmic
 divergences are canceled out.  Notice that there are always
 non-trivial solutions of the Pauli-Villars conditions equations
 (\ref{pvc}), e.g. ${\mathbf s}=(1,1,-2,-1), \,
 {\boldsymbol \mu^2}=(5,8,2,9)$. However, in the limit when the
 mass $M$ of the Pauli-Villars regulators goes to infinity we
 recover the quartic divergences of the vacuum energy-momentum
 tensor which now are of the form
  \beq
\langle T_{\mu\nu}\rangle_{\mbox {quart. div}}
&=& M^4 t^{(4)}_{\mu\nu}= \frac{c M^4}{2\pi^2 }
\left(
\begin{array}{cccc}
1  &   0  &  0  &  0   \\
0  & - 1  &  0  &  0   \\
0  &   0  & -1  &  0   \\
0  &   0  &  0  & -1   \\
\end{array}
\right),
\label{a33}
\eeq
 where $$c=  - \frac1{16}\sum_{i=1}^N s_i \mu_i^4 \log \mu_i$$ is an
 arbitrary constant given in terms of the regulating Pauli-Villars
 parameters. Notice that the sign of the vacuum energy correction
 might become  positive or negative depending of the choice of the
 regularization. In spite of the use of a non-covariant auxiliary
 cut-off the final result is covariant \cite{af}. However, the
 existence of an  ambiguity in the leading quartic divergence and
 its sign is a puzzling characteristic of quantum vacuum. From a
 renormalization viewpoint the ambiguities can be traced back to
 the locality of the cosmological constant term of the effective
 action. The effective value of the coupling has to be fixed by an
 explicit choice of  renormalization prescription.

 In the case of sub-leading divergences something similar occurs. The
 quadratic divergences  also acquire a covariant form in  Pauli-Villars
 regularization
 \beq
\langle T_{\mu\nu}\rangle_{\mbox {quad. div}}
&=& M^2 t^{(2)}_{\mu\nu}=\frac{c' M^2 m^2 }{2\pi^2 }
\left(
\begin{array}{cccc}
1  &   0  &  0  &  0   \\
0  & - 1  &  0  &  0   \\
0  &   0  & -1  &  0   \\
0  &   0  &  0  & -1   \\
\end{array}
\right),
\label{a3322}
\eeq
with
\beq
c'=  - \frac1{8}\sum_{i=1}^N s_i \mu_i^2 \log \mu_i.
\eeq
In the case of quadratic divergences there is one more specific
ambiguity, especially when they are calculated in the special
cosmological background depending on the Hubble parameter $H$.
Imagine we have obtained the result in the form
\beq
\langle T_{\mu\nu}\rangle
&=& M^4 t^{(4)}_{\mu\nu} + M^2 H^2 t^{(2)}_{\mu\nu} + \,...\,.
\label{a4}
\eeq
Now, in this expression $H$ is effectively used as a constant,
and therefore we can redefine the cut-off as
\ $M^2 \to {M^\prime}^2 = M^2 + \la H^2$, where $\la$ is
an arbitrary dimensionless parameter. As a result we arrive at
the new form of the power-like divergence,
\beq
\langle T_{\mu\nu}\rangle
&=& M^4 t^{(4)}_{\mu\nu} + M^2 H^2
\big[ t^{(2)}_{\mu\nu} + 2t^{(4)}_{\mu\nu} \big]
+ \,...\,.
\label{a5}
\eeq
with even greater degree of ambiguity. In the case of a theory
of the quantum field with mass $m$ one can perform a more
general redefinition
\ $M^2 \to {M^\prime}^2 = M^2 + \la H^2 + \tau m^2$,
with even more ambiguity, etc. It is important that the
logarithmic divergences are not affected by this ambiguity and,
in general, represent the most universal and well-defined part
of quantum corrections \cite{Salam-51} (see also \cite{MA-Eng}
for a recent discussion of the subject and further references).

One simple way to get free of the mentioned ambiguities is
by using the effective action method. The prescription which we
have already described above is simple. First one has to derive
the divergent and finite (at the level which is possible) of
effective action, add counterterms, perform renormalization.
At the second stage it is necessary to take a variational
derivative with respect to the metric (\ref{Tmn}) and obtain
the divergent part and/or renormalized EMT. In the next section
we shall see that this procedure works even in the situation with
SSB, where the procedure described above is essentially more
complicated than in the free field case.

%%%%%%%%%%%%%%%%%%%%%%%%%%%%%%%%%%%%%%%%%%%%%%%%%%%%%%%%%%%%%%
%%%%%%%%%%%%%%%%%%%%%%%%%%%%%%%%%%%%%%%%%%%%%%%%%%%%%%%%%%%%%%
%%%%%%%%%%%%%%%%%%%%%%%%%%%%%%%%%%%%%%%%%%%%%%%%%%%%%%%%%%%%%%
\section{  %% Spontaneous symmetry breaking
SSB and EMT in curved space}
%%%%%%%%%%%%%%%%%%%%%%%%%%%%%%%%%%%%%%%%%%%%%%%%%%%%%%%%%%%%%%

We start following Ref. \cite{Sponta}.
However, since our purpose is to consider the most simple
model with spontaneous symmetry breaking (SSB) in curved
space, we will consider the single real scalar field, while
in the mentioned reference the charged scalar was used.
The classical action of the field $\phi$ with a non-minimal
coupling and a self-interaction is
\beq
S_{sc}  &=&
\int d^4x\sqrt{-g}\,\Big\{\frac{1}{2}\,g^{\mu\nu}
\pa_{\mu}\phi\pa_{\nu}\phi
\,+\, \frac{1}{2}\,m^2\phi^2
\,+\,\frac{1}{2}\,\xi R\phi^2
\,-\,\frac{\lambda}{4!}\phi^4\Big\}\,.
\label{sinna}
\eeq
The dynamical equation for $\phi$ has the form
\beq
-\Box\phi +m^2\phi+\xi R\phi-\frac{1}{6}\lambda\phi^3\,=\,0\,.
\label{esatr}
\eeq
Consequently, the vacuum expectation value (VEV) for the
scalar field is then defined as solution of the equation
\beq
-\Box v +m^2v+\xi Rv-\frac{1}{6}\lambda v^3=0\,.
\label{ok}
\eeq
It is easy to see that there is no constant solution for this
equation for $\xi\neq 0$, while the value $\xi=0$ is inconsistent
with renormalizability of the theory (see, e.g., \cite{book}).
Hence we can find the solution for the vacuum expectation value
$v$ only in the form of the power series in $\xi R$,
\beq
v(x)=v_0+v_1(x)+v_2(x)+...\,.
\label{cham}
\eeq
In the zero-order approximation we meet the conventional flat-space
expression,
\beq
v_0^2 &=& \frac{6m^2}{\lambda}\,.
\label{saalleaas}
\eeq
As we have already mentioned above, in this paper we will use
the approximation of small curvature and are interested in the
first-order approximation only. At this level one can easily
find a non-local expression
\beq
v_1=\frac{\xi\,v_0}{\Box+\lambda v_0^2/3}\,R\,.
\label{fred}
\eeq
In a similar way, it is possible to construct further approximations,
but this is beyond the scope of the present paper. Thus, let us
concentrate on the expression (\ref{fred}) and simplify it further
by neglecting  terms with derivatives of the scalar curvature. Of
course, this approximation works only for the sufficiently large
value of the square of the physical mass of the scalar excitation
near the point of the minimum, $2m^2=\lambda v_0^2/3$. Then we
arrive at the quantity
\beq
v_1\,\approx\, \frac{3\xi}{\lambda v_0}\,R\,.
\label{fred-red}
\eeq
It is clear that the same solution can be obtained directly
from Eq. (\ref{ok}) if we disregard the term $\Box v$ and use
(\ref{saalleaas}). In our opinion the approach followed here
is better, because it enables one to control the approximation.
In further calculations we shall use the expression for classical
solution of the theory in the point of the minima of the SSB
problem,
\beq
\phi_{0c} &=& v \,=\,v_0 + v_1\,,
\quad \mbox{where}\quad
v_0^2 = \frac{6m^2}{\lambda}
\quad \mbox{and}\quad
v_1 = \frac{3\xi}{\lambda v_0}\,R\,.
\label{fred-green}
\eeq

The renormalization of the vacuum sector of the theory with
SSB has been described in great detail in \cite{Sponta}, thus,
we shall not elaborate on it  here. Instead, let us discuss
the definition of
the EMT at quantum level for the theory with SSB. The classical
energy-momentum tensor of the field $\phi$ in the external
metric field $g=g_{\mu\nu}$ is defined by the relation
\beq
T_{\mu\nu} &=& - \frac{2}{\sqrt{-g}}\,g_{\mu\al}\,g_{\nu\be}\,
\frac{\delta S[g,\phi]}{\de g_{\al\be}}\,.
\label{caza}
\eeq
At quantum level, the energy-momentum tensor is given by
\beq
\langle T_{\mu\nu}\rangle
\,=\,-\,\frac{2}{\sqrt{-g}}\,
g_{\mu\alpha}g_{\nu\beta}\,\Bigl\langle 0\Big|\frac{\delta
S[g,\hat{\phi}]}{\delta g_{\alpha\beta}}\Big|0 \Bigr\rangle\,,
\eeq
where \ $\hat{\phi}$ \ is quantized field, \
$\hat{\phi}\sim u\hat{a}^{\dagger}+u^{*}\hat{a}$ \ and \
$\hat{a}\,|0\rangle=0$.
As far as $g_{\mu\al}$ is classical external field, so we can
take it out of the $\langle|..|\rangle$ freely.

According to our previous discussion, we will follow the functional
representation of Quantum Field Theory, where the basic object is
the generating functional of vertex function, or effective action,
$\Ga=\Ga[g,\phi]$. For the case of a scalar field it is
defined as a solution of the functional equation (see, e.g.,
\cite{book} for a introduction)
\beq
\exp \left\{{\frac{i}{\hbar}\Gamma[g,\phi]}\right\}
&=&
\int d\bar{\phi} \,\exp{\Big\{
\frac{i}{\hbar}\Big(S[g,\bar{\phi}+\phi]
- \frac{\de\Ga[g,\phi]}{\de\phi}\,\bar{\phi}\Big]\Big\}}.
\label{eq4}
\eeq

In this work we restrict consideration by the one-loop approximation,
when the effective action in Eq. (\ref{Tmn}) becomes the sum of the
classical term and of the one-loop correction
\beq
\Ga^{(1)}[\phi,\,g_{\mu\nu}]
&=&
S[\phi,\,g_{\mu\nu}] \, + \,\hbar{\bar \Ga}^{(1)}[\phi,\,g_{\mu\nu}]\,.
\label{EA-1}
\eeq
Then the one-loop EMT of the vacuum can be cast into the form
\beq
\langle T_{\mu\nu}(x) \rangle^{(1)}
&=&
T_{\mu\nu}(x) + {\bar T}_{\mu\nu}^{(1)}(x)\,,
\label{Tmn-1}
\eeq
where the first term is classical contribution,
\beq
T_{\mu\nu} &=&
-\,\frac{2}{\sqrt{-g}}\,g_{\mu\al}\,g_{\nu\be}
\,\frac{\de S}{\de g_{\al\be}}\Big|_{\phi \to \phi_0}\,.
\label{Tmn-2}
\eeq
and the second one is one-loop correction to it,
\beq
{\bar T}_{\mu\nu}^{(1)} &=&
-\,\frac{2\,\hbar}{\sqrt{-g}}\,g_{\mu\al}\,g_{\nu\be}
\,\frac{\de {\bar \Ga}^{(1)}}{\de g_{\al\be}}
\Big|_{\phi \to \phi_0}\,.
\label{Tmn-3}
\eeq
In both cases $\phi_0$ is the solution of the equations of
motion. If we deal with purely classical theory, then one has
to replace in (\ref{Tmn-2}) the value $\phi_0 = \phi_{0c}$
from Eq. (\ref{fred-green}). After the one-loop correction
is taken into account, we have
\beq
\frac{\de S[g,\phi_0]}{\de\phi}
\,+\,\hbar\,\frac{\de\bar{\Ga}^{(1)}[g,\phi_0]}{\de\phi}\,=\,0\,,
\label{andiamo}
\eeq
where the replacement $\phi \to \phi_0$ should be performed
after variational derivative. The Eq. (\ref{andiamo}) can be
solved by iterations in $\hbar$. At one-loop level
\beq
\phi_0 &=& \phi_{0c} + \hbar\phi_1\,,
\label{phi1}
\eeq
where $\phi_{0c}$ is the classical solution (\ref{fred-green}).
In order to find $\phi_1$ one has to replace (\ref{phi1}) into
(\ref{andiamo}). In the first order in $\hbar$ we meet the equation
\beq
\frac{\delta^2 S[g,\phi_{0c}]}{\delta\phi\delta\phi}\phi_1
+ \frac{\delta \bar{\Gamma}^{(1)}[g,\phi_{0c}]}{\delta\phi}=0\,,
\eeq
and obtain the solution in the form
\beq
\phi_1 &=& -\,\Biggl( \frac{\delta^2
S[g,\phi_{0c}]}{\de\phi\,\de\phi}\Biggl)^{-1}
\,\frac{\de\,\bar{\Ga}^{(1)}[g,\phi_{0c}]}{\delta\phi}
\label{phi2}
\eeq
and, therefore,
\beq
\phi_0 &=& \phi_{0c}\,-\,\hbar\,\Biggl( \frac{\de^2
S[g,\phi_{0c}]}{\de\phi\,\de\phi}\Biggl)^{-1}
\,\frac{\de\,\bar{\Ga}^{(1)}[g,\phi_{0c}]}{\de\phi}\,.
\label{wills}
\eeq
One has to replace this formula into the expression for EMT,
\beq
\langle T_{\mu\nu} \rangle
&=& -\, \frac{2}{\sqrt{-g}}\,g_{\mu\al}g_{\nu\be}
\,\frac{\de}{\de g_{\al\be}}
\Big\{ S[g,\phi_0]+\hbar\,\bar{\Ga}^{(1)}[g,\phi_0]\Big\}\,.
\eeq
In this way we arrive at the general expression for the EMT
in the scalar theory with SSB,
\beq
\langle T_{\mu\nu}\rangle
&=& -\,\frac{2}{\sqrt{-g}}\,g_{\mu\al}\,g_{\nu\be}\,
\Biggl\{
\frac{\delta\,
S[g,\phi_{0c}]}{\de g_{\al\be}}
\nonumber
\\
&+&
\hbar\,\frac{\de\,
\bar{\Ga}^{(1)}[g,\phi_{0c}]}{\de g_{\al\be}}
\,-\,\hbar\,\frac{\de^2 S[g,\phi_{0c}]}{\de g_{\al\be} \de\phi}
\Biggl(\frac{\delta^2 S[g,\phi_{0c}]}{\de\phi\,\de\phi}\Biggr)^{-1}
\,\frac{\de\,\bar{\Ga}^{(1)}[g,\phi_{0c}]}{\de\phi}
\Biggr\}\,.
\label{Lvr}
\eeq
The first term inside the brackets is classical, the second is
typical for the free theory and actually does not depend too much
on the kind of such theory. The last term emerges only due to the
fact that we deal with the interacting theory. In the free theory
this term is zero.

It proves useful to define
\beq
\langle T_{\mu\nu}\rangle\,=\,\langle T_{\mu\nu}\rangle_{v}
\,+\,\langle T_{\mu\nu}\rangle_{i},
\label{lavo}
\eeq
where
\beq
\langle T_{\mu\nu}\rangle_{v}
\,=\,-\,\frac{2}{\sqrt{-g}}\,g_{\mu\alpha}\,g_{\nu\beta}
\Biggl\{\frac{\delta\,
S[g,\phi_{0c}]}{\delta g_{\alpha\beta}}+\hbar\frac{\delta\,
\bar{\Gamma}^{(1)}[g,\phi_{0c}]}{\delta
g_{\alpha\beta}}\Biggr\}
\label{lavo1}
\eeq
and
\beq
\langle T_{\mu\nu}\rangle_{i}
\,=\,\frac{2\hbar}{\sqrt{-g}}\,g_{\mu\alpha}g_{\nu\beta}\,
\Biggl\{
\frac{\delta^2
S[g,\phi_{0c}]}{\de g_{\al\be}\de\phi}
\,\Biggl( \frac{\de^2 S[g,\phi_{0c}]}{\de\phi\,\de\phi}\Biggr)^{-1}
\,\frac{\de\,\bar{\Ga}^{(1)}[g,\phi_{0c}]}{\de\phi}
\Biggr\}\,.
\label{shkfe}
\eeq
The quantities \ $\langle T_{\mu\nu}\rangle_{v}$ \ and \
$\langle T_{\mu\nu}\rangle_{i}$ \ represent the vacuum and
induced parts of the EMT, respectively. Both quantities will
be calculated in this work in the context of SSB.

%%%%%%%%%%%%%%%%%%%%%%%%%%%%%%%%%%%%%%%%%%%%%%%%%%%%%%%%%%%%%%
%%%%%%%%%%%%%%%%%%%%%%%%%%%%%%%%%%%%%%%%%%%%%%%%%%%%%%%%%%%%%%
%%%%%%%%%%%%%%%%%%%%%%%%%%%%%%%%%%%%%%%%%%%%%%%%%%%%%%%%%%%%%%
\section{Covariance and conservation of vacuum EMT}
%% energy-momentum tensor
%%%%%%%%%%%%%%%%%%%%%%%%%%%%%%%%%%%%%%%%%%%%%%%%%%%%%%%%%%%%%%

The conservation of the energy-momentum tensor of vacuum is
always regarded as the main requirement for the consistency of
the theory (see, e.g., \cite{birdav} and further references therein).
As far as we deal with the lower-derivative approximation, the
condition of conservation $\na^\mu\langle T_{\mu\nu}\rangle=0$,
together with the requirement that the EMT should be derived
as variational derivative of covariant effective action, can
fix the algebraic form of $\langle T_{\mu\nu}\rangle$ completely,
leaving the room for only two numerical parameters in case
of quadratic and logarithmic divergences and finite part and
for a single numerical parameter for the quartic divergences
case.

The reasons for this special importance of the conservation law
are as follows. For the divergent parts of effective action the
situation is especially simple, because we know it should be
local (see \cite{Ren-Curved} for a recent discussion of this
issue in curved space-time). As we have already explained in
the previous section, in the lower-derivative sector this
means that the possible counterterms have the form of the
Einstein-Hilbert term and of the cosmological constant term
(\ref{EH}). Consequently, the divergent part of the vacuum EMT
should consist of only two structures, namely
\beq
\langle T_{\mu\nu}\rangle
\,=\, C_1\,g_{\mu\nu}\,+\,C_2\,G_{\mu\nu}\,,
\label{divss}
\eeq
where $C_1=k_4\Om^4+k_2\Om^2+k_L\ln(\Om/\mu_0)$ and
$C_2=l_2\Om^2+l_L\ln(\Om/\mu_0)$, with \ $k_4,\,k_2,\,k_L$
\ and \ $l_2,\,l_L$ being numerical constants. The values
of these constants depend on the choice of the quantum theory,
on the order of loop expansion, but the structure of divergent
part must be always like in (\ref{divss}).

Concerning the finite part of the EMT of the vacuum, it is
possible to have much more complicated expression than the one
presented in (\ref{divss}), as a result of resummation of the
series in curvatures and Green functions \cite{DCCR}. One can
have an indication to this possibility, e.g., from the
calculation in conformal variables \cite{Shocom}. However,
as far as we are going to perform a relatively simple
calculation in the ${\cal O}(R)$-approximation, there is
no room for non-localities in the effective action, so what
one should expect as a  result is the same expression
(\ref{divss}).

Needless to say that  (\ref{divss}) is the only form of EMT
which can be derived from some action principle and also is
the only form which satisfies conservation law. Let us start
from a well-known derivation of this relation in general case
and then consider the same thing in view of Eq. (\ref{Lvr}).

The effective action $\Ga$ is covariant scalar functional
depending on metric $g_{\mu\nu}$ and scalar field $\phi$.
If we perform infinitesimal general coordinate transformation
\ $x^\al \to {x^\prime}^\al = x^\al + \xi^\al(x)$, these
two fields transform according to the known rules
\beq
\de g_{\mu\nu} = - \na_\mu\xi_\nu  - \na_\nu\xi_\mu\,,
\qquad
\de \phi = - \,\xi^\mu \pa_\mu\phi\,.
\label{trans}
\eeq
Then the identity corresponding to diffeomorphism invariance
of $\Ga$ is
\beq
\int d^4x\sqrt{-g}\,
\Biggl\{
\frac{2}{\sqrt{-g}}\,
\frac{\delta \Ga[g,\phi]}{\de g_{\mu\nu}}\,\na_\mu\xi_\nu
\,+\,\frac{1}{\sqrt{-g}}\,
\frac{\de \Ga[g,\phi]}{\de \phi}\,\xi^\mu \na_\mu\phi
\Biggl\}\,=\,0\,.
\label{our}
\eeq
Now we take into account that the functional derivative
vanish on-shell, that means
\beq
\frac{\de \Ga[g,\phi_0]}{\de \phi}\,=\,0\,.
\label{cons-1}
\eeq
Then integrating the first term in (\ref{our}) and taking into
account the definition of EMT (\ref{Tmn}), we arrive at the
conservation law, \ $\na^\mu\langle T_{\mu\nu}\rangle_{\phi_0}=0$.

Let us now see how the same considerations look when we perform
the expansion of $\Ga[g,\phi_0]$ into series in $\hbar$. Here we
are interested in the expansion up to the first order and the
main question is whether the two parts EMT, namely vacuum
and induced ones, (\ref{shkfe}) and (\ref{lavo1}), do satisfy
the conservation law separately or only when they are summed up.

At zero order everything is quite obvious, for we have
\beq
\frac{\de S[g,\phi_{0c}]}{\de \phi}\,=\,0\,,
\qquad
\mbox{and}
\qquad
\na^\mu T_{\mu\nu}\big|_{\phi_{0c}}\,=\,0\,.
\label{cons-2}
\eeq

At the first order in $\hbar$ we notice that the conservation
law is satisfied only on-shell. Now, the solution
\ $\phi_0 = \phi_{0c} + \hbar\phi_1$ \ of (\ref{phi2}), was found
exactly to provide that $\phi_0$ is the solution of the effective
equations of motion at one loop. Hence, we should expect that
neither one of the two terms  (\ref{shkfe})
abd (\ref{lavo1}) will satisfy the conservation equation and
only for their sum this equation must be valid,
\beq
\na^\mu\langle T_{\mu\nu}(\phi_{0c})\rangle_v
\,+\,\na^\mu\langle T_{\mu\nu}(\phi_{0c})\rangle_i \,=\,0\,.
\label{lavo-cons}
\eeq
On the other hand, this means that the sum (\ref{lavo}) should
have the form (\ref{divss}) while each term can have more
arbitrary form, for example the Ricci tensor and scalar
curvature term may not form Einstein tensor.

Finally, both the covariance arguments and conservation law
indicate that the quantum EMT of vacuum, in the low-energy
sector of the theory, must have the form (\ref{divss}) even
in the presence of SSB which produce much more sophisticated
forms of EMT, such as (\ref{Lvr}). The restricted form of
the vacuum EMT (\ref{divss}) should hold even at higher
loop orders, at least for divergent
contributions\footnote{For the
finite part, it is still  possible to have  a much more
complicated form of EMT due to the possible resummation
in the expansions in curvature tensor, as it was discussed
in \cite{DCCR}.}.
This is a strong statement and it is worthwhile to check it
by direct calculation, at least in the one-loop order. We shall
do it in the next section.

%%%%%%%%%%%%%%%%%%%%%%%%%%%%%%%%%%%%%%%%%%%%%%%%%%%%%%%%%%%%%%
%%%%%%%%%%%%%%%%%%%%%%%%%%%%%%%%%%%%%%%%%%%%%%%%%%%%%%%%%%%%%%
%%%%%%%%%%%%%%%%%%%%%%%%%%%%%%%%%%%%%%%%%%%%%%%%%%%%%%%%%%%%%%
\section{Derivation of stress tensor: classical part}
%%%%%%%%%%%%%%%%%%%%%%%%%%%%%%%%%%%%%%%%%%%%%%%%%%%%%%%%%%%%%%

In this section we shall derive the EMT of vacuum at
classical level and then, in the next sections, turn to
the one-loop contributions.

Let us perform the calculation of the EMT $\,T_{\mu\nu}\,$ in
the vacuum state, which is characterized by the VEV of scalar
field defined in Eq. (\ref{fred-green}). The calculation of
$\,T_{\mu\nu}\,$ is rather trivial and we obtain
\beq
T_{\mu\nu}
&=&
-\,\frac{2}{\sqrt{-g}}\,g_{\mu\alpha}g_{\nu\beta}\,
\frac{\de\,S[g,\phi]}{\de g_{\al\be}}
\,=\,\Big(2\xi-\frac{1}{2}\Big)g_{\mu\nu}(\nabla\phi)^2
+(1-2\xi)(\partial_{\mu}\phi)(\partial_{\nu}\phi)
\nonumber
\\
&+&
2\,\xi\phi(g_{\mu\nu}\Box\phi-\nabla_{\mu}\nabla_{\nu}\phi)
+\xi\phi^2\Big(R_{\mu\nu}-\frac{1}{2}Rg_{\mu\nu}\Big)
+\frac{1}{2}g_{\mu\nu}m^2\phi^2
+\frac{\lambda}{24}g_{\mu\nu}\phi^4\,.
\label{timbos}
\eeq
The trace of the scalar EMT on-shell (\ref{esatr}) can be
easily reduced to the form
\beq
T^{\mu}_{\mu}&=&
(6\xi-1)\Big[(\na\phi)^2 + R\phi^2 - \frac{\la}{6}\,\phi^4\Big]
\,+\, 2(1-3\xi)\,m^2\phi^2\,.
\label{trace}
\eeq
We observe that for $m^2=0$ and $\xi=1/6$ we have $T^{\mu}_{\mu}=0$.
However, we are interested in the massive case given by Eq.
(\ref{fred-green}). At this point it is worthwhile to discuss
the practical realization of the ${\cal O}(R)$ approximation,
which we will follow in this section. The main question what to
do with the derivatives of $\phi_{0c}=v_0+v_1$. Since $v_0$ is a
constant, its derivative is obviously zero. Furthermore, a
derivative of $v_1$ gives us
\beq
\na_\al\phi_{0c} &=&
\na_\al v_1 = \frac{3\xi}{\lambda v_0}\,\na_\al R
\label{deri v1}
\eeq
and, therefore, goes beyond the limits of our approximation.
As a result we can always treat $R$ and $v_1$ as constants,
that leads to great simplification of all calculation.

Replacing (\ref{fred-green}) into (\ref{timbos}) and keeping only
terms linear in curvature tensors, after small algebra we arrive at
\beq
T_{\mu\nu}(\phi_{0c})
&=&
\xi v_0^2\,\Big(R_{\mu\nu} - \frac12\,Rg_{\mu\nu}\Big)
\,-\, \frac{\la v_0^4}{12}\,g_{\mu\nu}\,.
\label{classic}
\eeq
This expression is nothing else but the induced contribution to
the Einstein equations. It is natural to attribute it to the
gravitational part of these equations, which can be, eventually,
written as
\beq
 \Big(\frac{1}{8\pi G_{vac}} + \frac{1}{8\pi G_{ind}}\Big)
\Big(R_{\mu\nu} - \frac12\,Rg_{\mu\nu}\Big)
\,-\, \Big(\rho_\La^{vac} \,+\, \rho_\La^{ind}\Big) \,g_{\mu\nu}
\,=\,T_{\mu\nu}^{\rm matter}\,,
\label{Ein-1}
\eeq
where
\beq
\frac{1}{8\pi G_{ind}} = - \xi v_0^2
\qquad
\mbox{and} \qquad
\rho_\La^{ind} = -\frac{\la v_0^4}{12}\,.
\label{Ein-2}
\eeq
In this equation $\,G_{vac}\,$ and $\,\rho_\La^{vac}\,$ denote the
vacuum Newton constant and the cosmological constant density,
which are independent parameters that are originally present
in the action of
the theory. Contrary to that, $\,G_{ind}\,$ and $\,\rho_\La^{ind}\,$
are induced quantities which depend on the details of the quantum
theory of matter fields under consideration.

The induced contributions here are due to the SSB, and an
equivalent mechanism of their generation is working also for
the Standard Model (SM) and Grand Unification Theories (GUT's).
The values of induced and vacuum
cosmological constants are known to be, at least, 55 orders of
magnitude greater than their sum in (\ref{Ein-1}), that gives
rise to the cosmological constant problem \cite{weinberg89}
(see also references therein and \cite{CC-nova} in relation to
renormalization of the cosmological constant). On the contrary
 the relative magnitude of $\,G_{ind}\,$, namely
\beq
\frac{G_{ind}}{G_{vac}}\,=\,-\,\frac{8\pi \xi v_0^2}{M_P^2}\,,
\label{relative G}
\eeq
is small for the SM case where $v_0^2 \approx 10^5 GeV^2$.
Even if the value of $\xi$ corresponds to the Higgs inflation,
$\xi \approx 4\times 10^4$, the Planck suppression is strong
due to the relatively huge value $M_P^2 \approx 10^{38}GeV^2$
and hence the contribution of (\ref{relative G}) is irrelevant.

However, the situation can be quite different in GUT's, where (in
the supersymmetric versions) we have $\,v_0^2 \approx 10^{32} GeV^2$.
Then for the mentioned above magnitude of $\xi$ we arrive at the
estimate $G_i/G_v \approx 1$ and the effective sum in (\ref{Ein-1})
becomes close to zero. This means the value of $1/G_v$ must be taken
about twice larger than the observed sum. Hence, the classical
screening due to induced value may be relevant in this case. In
the rest of this paper we shall check that the quantum effects
do not break the structure of (\ref{Ein-1}) and calculate quantum
corrections to the quantities $\,G_{ind}\,$ and $\,\rho_\La^{ind}\,$
in (\ref{Ein-2}).

%%%%%%%%%%%%%%%%%%%%%%%%%%%%%%%%%%%%%%%%%%%%%%%%%%%%%%%%%%%
%%%%%%%%%%%%%%%%%%%%%%%%%%%%%%%%%%%%%%%%%%%%%%%%%%%%%%%%%%%
%%%%%%%%%%%%%%%%%%%%%%%%%%%%%%%%%%%%%%%%%%%%%%%%%%%%%%%%%%%
\section{One-loop calculation in the
${\cal O}(R)$-approximation}
%%%%%%%%%%%%%%%%%%%%%%%%%%%%%%%%%%%%%%%%%%%%%%%%%%%%%%%%%%%%%%

Let us now perform quantum calculations using the expressions
for vacuum and induced parts, (\ref{shkfe}) and (\ref{lavo1}).
The calculations will be done in the local momentum
representation and covariant momentum cut-off regularization.
For better organization, this section is divided into subsections.
First, we consider some general notions, then derive the flat-space
result, then present some minimal mathematical tools for the local
momentum representation, and finally perform derivation of the more
complicated, curvature-dependent part.

%%%%%%%%%%%%%%%%%%%%%%%%%%%%%%%%%%%%%%%%%%%%%%%%%%%%%%%%%%%%%%%%%%%%%
\subsection{General considerations and derivation of
$\langle T_{\mu\nu}\rangle_{v}$}

Our starting point will be the one-loop effective action,
${\bar \Ga}^{(1)}[g,\phi]$. By construction, this is the
effective action in the theory with unbroken symmetry.
One can write, using derivative expansion,
\beq
{\bar \Ga}^{(1)}[g,\phi] =
\int d^4x\sqrt{-g} \Big\{ - {\bar V}_{eff}(\phi)
+ \frac{1}{2}\na_{\mu}\phi\cdot k_{\phi}\Big(\frac{\Box}{m^2}\Big)
\na^{\mu}\phi
+ \frac{1}{2} \phi^2 k_{\xi}\Big(\frac{\Box}{m^2}\Big)R
\,+...\Big\}\,,
\label{popots}
\eeq
where the effective potential part has the form
\beq
 {\bar V}_{eff}(\phi) &=& V_0 + V_1\,R + {\cal{O}}(R_{...}^2)\,,
\label{EP}
\eeq
which was recently calculated using covariant momentum cut-off
in \cite{CorPot} and $k_{\phi}\Big(\frac{\Box}{m^2}\Big)$
and $k_{\xi}\Big(\frac{\Box}{m^2}\Big)$ are the form factors
which also contain different powers of derivatives.
The expansion in (\ref{popots}) is infinite, but we can easily
set the limit on it, following the same approach which was used
in the previous section. For $\xi=0$ we know $\phi_{0c}=const$,
according to Eq. (\ref{esatr}). Therefore, any derivatives of
$\phi_{0c}$ are actually proportional to $\xi$ and hence to $R$.

As far as we are interested only in ${\cal{O}}(R)$\,- terms, we
can take only constant part of the form-factor $k_{\xi}$ in
(\ref{popots}), and also strongly restrict $k_{\phi}$ form
factor, also by taking its constant part. Hence we can trade
\beq
k_{\phi}\Big(\frac{\Box}{m^2}\Big)\rightarrow Z(\phi)\,,
\quad
\mbox{and}
\quad
k_{\xi}\Big(\frac{\Box}{m^2}\Big)\rightarrow \chi(\phi)\,.
\label{Z}
\eeq
Futhermore, in the given approximation the term
$\frac{1}{2}\phi^2\,\chi(\phi)\,R$ is a part of the effective
potential $V_1=V_1(\phi)$. So, for us $\bar{\Gamma}^{(1)}[g,\phi]$
becomes
\beq
\bar{\Gamma}^{(1)}(g,\phi)=\int d^4x\sqrt{-g}\,\Big\{\,
- {\bar V}_{eff}(\phi)
+\frac{1}{2}Z(\phi)(\na\phi)^2\Big\}\,,
\eeq
with \ ${\bar V}_{eff} = V_0(\phi)+V_1(\phi)\,R$ \ and
\ $\phi\rightarrow\phi_{0c}$.

Let us consider
\beq
\int d^4x\sqrt{-g}\, Z(\phi)(\na\phi)^2
&=& \int d^4x\sqrt{-g}\, \na_\mu\chi^\mu
-\int d^4x\sqrt{-g}\, Z(\phi)\,\phi\,\Box\phi
\nonumber
\\
&-& \int d^4x\sqrt{-g}\, Z\,'(\phi)(\na\phi)^2.
\eeq
For \ $\phi\rightarrow\phi_{0c}$, the quantity $\Box\phi$ can be
written as
\beq
\Box\phi_{0c} = \Box (v_0+v_1)
= \frac{\xi v_0}{\Box+2m^2}\,\Box R
= \frac{\xi v_0}{2m^2}\,\Box R \,+\,{\cal O}(\Box^2 R)\,.
\label{A1}
\eeq
On the other hand,
\beq
\na_{\mu}\phi_{0c}=\na_{\mu} v_0+\na_{\mu} v_1
=\na_{\mu} v_1
= \frac{\xi v_0}{2m^2} \na_{\mu} R
\,+\,{\cal O}(\na^3 R)\,.
\label{A2}
\eeq
It is now obvious that, because of \ $\Box \ll m^2$ \ for
\ $\phi_{0c}$, the whole quantity \ $(\na\phi)^2$ \ is beyond
our approximation \ ${\cal O}(R)$. Finally, we can restrict our
consideration by the effective potential, using the expression
\beq
\bar{\Gamma}^{(1)}[g,\phi_{0c}]
&=&
-\,\int d^4x\sqrt{-g}\,{\bar V}_{eff}(\phi_{0c})\,.
\label{GV}
\eeq
The renormalized expression of the potential is \cite{CorPot}
\beq
&& {\bar V}_{eff}^{ren}(g_{\mu\nu},\,\ph)
\,=\,
V_0^{ren} \,+\,V_1^{ren}R
\label{REN-potya}
\\
&& =\,\frac{1}{2(4\pi)^2}
\Big[\frac12 \big(V^{\prime\prime}-m^2\big)^2
- \Big(\xi-\frac16\Big)\, R\,
\big(V^{\prime\prime}-m^2\big)\Big]
\ln \Big(\frac{V^{\prime\prime}-m^2}{\mu^2}\Big)\,.
\nonumber
\eeq
In this expression we used a general form of classical interaction
term $V=V(\ph)$, but later on it will be replaced by $V=\la\ph^4/4\!$.

For the sake of completeness we will also consider the divergent
part of the non-renormalized potential, in the local momentum
cut-off regularization. In the given approximation we have
\ ${\bar V}_{eff}^{div}(g_{\mu\nu},\,\ph)
\,=\, V_0^{div} \,+\,V_1^{div}R$, \ where
\beq
\bar{V}_0^{div}
&=&   \frac{1}{32\pi^2}\,\Big\{
\Om^2V^{\prime\prime}
- \frac12\,\big(V^{\prime\prime}-m^2\big)^2
\ln \frac{\Om^2}{m^2}\Big\}\,,
\label{potya-div}
\\
\bar{V}^{div}_1
&=&  \frac{1}{32\pi^2}\,\Big(\xi-\frac16\Big)\,
\Big\{\,-\,\Om^2\,+\,\big(V^{\prime\prime}-m^2\big)
\,\,\ln \frac{\Om^2}{m^2}\Big\}\,,
\label{scalar div}
\eeq

It proves useful to introduce a notation for the one-loop
contributions to the equations of motion for a scalar field,
\beq
{\bar \vp}^{(1)} &=& {\bar \vp}^{(1)}_{div}
+ {\bar \vp}^{(1)}_{fin}
%% \nonumber \\
\,=\, \frac{1}{\sqrt{-g}}\,
\frac{\de {\bar \Ga}^{(1)}}{\de \phi}\Big|_{\phi_{0c}}
\,=\,-\,\frac{\pa {\bar V}_{eff}^{(1)}}{\pa \phi}\Big|_{\phi_{0c}}\,.
\label{eps}
\eeq
After adding the corresponding counterterm, we will also have
\ ${\bar \vp}^{(1)}_{ren}$. Let us now remember that
\ ${\bar V}_{eff} = V_0(\phi)+V_1(\phi)\,R$, \
$\phi_{0c}=v_0+v_1$, and define also
\beq
{\bar \vp}^{(1)} &=&
-\,\frac{\pa {\bar V}_0^{(1)}}{\pa \phi}\Big|_{\phi_{0c}}
-\,R\,\frac{\pa {\bar V}_1^{(1)}}{\pa \phi}\Big|_{\phi_{0c}}
\nonumber
\\
&=&
-\,\frac{\pa {\bar V}_0^{(1)}}{\pa \phi}\Big|_{v_0}
\,-\,\frac{\pa^2 {\bar V}_0^{(1)}}{\pa \phi^2}\Big|_{v_0}v_1
\,-\,R\,\frac{\pa {\bar V}_1^{(1)}}{\pa \phi}\Big|_{v_0}
\,=\,{\bar \vp}^{(1)}_0 \,+\,{\bar \vp}^{(1)}_1\,.
\label{eps-12}
\eeq
Obviously, both ${\bar \vp}^{(1)}_0$ and ${\bar \vp}^{(1)}_1$
have finite and divergent parts and after adding counterterms
we can also define their renormalized versions. In this paper
we will calculate divergent and renormalized quantities only,
but the original finite parts can be calculated in the same
way using, e.g., the effective potential from \cite{CorPot}.
The last relevant observation is that, within the ${\cal O}(R)$
approximation adopted here we can treat all versions of
${\bar \vp}^{(1)}_0$ and ${\bar \vp}^{(1)}_1$ (divergent, finite
non-renormalized, counterterms and renormalized) as constants.

Let us now calculate the simplest quantum term
$\langle T_{\mu\nu}\rangle_{v}$, defined in (\ref{lavo1}).
Starting from (\ref{GV}) we can easily arrive at
\beq
\langle {\bar T}_{\mu\nu}\rangle_{v}
&=& -\,\frac{2\,\hbar}{\sqrt{-g}}
\,g_{\mu\alpha}\,g_{\nu\beta}\,
\frac{\de\,{\bar \Ga}^{(1)}[g,\phi_{0c}]}{\de g_{\al\be}}
\nonumber
\\
&=& -\,2\,\hbar V_1(v_0)\Big(R_{\mu\nu} - \frac12\,Rg_{\mu\nu}\Big)
\,+\,\hbar V_0(v_0)\,g_{\mu\nu}
\,+\,\hbar \,v_1\,g_{\mu\nu}\,
\frac{\pa {\bar V}_0^{(1)}}{\pa \phi}\Big|_{v_0}
\nonumber
\\
&=& -\,2\,\hbar V_1(v_0)\,G_{\mu\nu}
\,+\,\hbar V_0(v_0)\,g_{\mu\nu}
\,-\,\frac{\hbar \,\xi\,v_0}{2m^2}\,R\,
{\bar \vp}^{(1)}_0\,g_{\mu\nu}\,.
\label{lavo10}
\eeq
This formula is remarkable, because it confirms what we have
anticipated in the previous section. The first two terms in the
last expression are quantum contributions to the Einstein tensor
and cosmological constant part in the Einstein equations. However,
the last term looks odd, for it violates covariance, conservation
law and can not be derived from the action principle. So, we
should hope that it will cancel with the corresponding contribution
from $\langle {\bar T}_{\mu\nu}\rangle_{i}$ in (\ref{shkfe}),
which is the last term in (\ref{Lvr}). Let us see whether this
really happens in the next section.

%%%%%%%%%%%%%%%%%%%%%%%%%%%%%%%%%%%%%%%%%%%%%%%%%%%%%%%%%%%%%%%%%%%%%
\subsection{Calculation of $\langle {\bar T}_{\mu\nu}\rangle_{i}$}

Our first step will be to rewrite the expression (\ref{shkfe}) for
$\langle {\bar T}_{\mu\nu}\rangle_{i}$ in a more useful and
detailed form
\beq
\langle T_{\mu\nu}(x)\rangle_{i}
&=&
2\hbar\,g_{\mu\alpha}(x)\,g_{\nu\beta}(x)\,
\int d^4y\sqrt{-g(y)} \int d^4z\sqrt{-g(z)}\,
\Biggl(\frac{1}{\sqrt{-g}}\,
\frac{\delta^2 S[g,\phi_{0c}]}{\de g_{\al\be}(x)\de\phi(y)}\Biggr)
\nonumber
\\
&\times&
\Biggl( \frac{1}{\sqrt{-g(y)}}\,
\frac{\de^2 S[g,\phi_{0c}]}{\de\phi(y)\,\de\phi(z)}\Biggr)^{-1}
\,\times\,
\Biggl(\frac{1}{\sqrt{-g(z)}}\,
\frac{\de\,\bar{\Ga}^{(1)}[g,\phi_{0c}]}{\de\phi(z)}
\Biggr)\,.
\label{ind-1}
\eeq
Let us note that the metric-dependent quantities are always
understood through the normal coordinate expansions (see some
details of this technique in Appendix A).

The next step is to derive all three factors inside the integrals
of the Eq. (\ref{ind-1}). We need to perform this calculation in the
${\cal O}(R)$ approximation, which we follow here. The first
factor can be obtaned by variating (\ref{timbos}) with respect to
$\phi$ or just taking a second variation of the action. After some
algebra we arrive at
\beq
\frac{1}{\sqrt{-g}}\frac{\de^2S}{\de\phi (y) \de g_{\mu\nu}(x)}
&=&
\xi\phi\big(\nabla_{\mu}\nabla_{\nu}-g_{\mu\nu}\Box\big)
+(2\xi-1)(\nabla_{\mu}\phi)\nabla_{\nu}\nonumber\\
&+&\Big(\frac{1}{2}
-2\xi\Big)g_{\mu\nu}(\nabla^{\lambda}\phi)\nabla_{\lambda}
+\xi\big(\nabla_{\mu}\nabla_{\nu}\phi-g_{\mu\nu}\Box\phi\big)\nonumber\\
&-&\xi\phi\Big(R_{\mu\nu}-\frac{1}{2}R\,g_{\mu\nu}\Big)
+\frac{1}{2}m^2\phi\,
g_{\mu\nu}-\frac{\lambda}{6}\phi^3g_{\mu\nu}\,.
\label{variat}
\eeq

Now one can replace in the last expression
$\phi \to \phi_{0c} = v_0 + v_1$ and remember that all
derivatives of $\phi_{0c}$ are beyond our approximation. In this
way we obtain
\beq
\frac{1}{\sqrt{-g}}\,g_{\al\mu}g_{\be\nu}\,
\frac{\de^2S}{\de g_{\al\be}\de\phi}\Big|_{\phi_{0c}}
&=&
\xi\phi_{0c}\,\big(\nabla_{\mu}\nabla_{\nu}-g_{\mu\nu}\Box\big)
\,-\,\xi\phi_{0c}\,\Big( R_{\mu\nu} -\frac{1}{2}\,R\,g_{\mu\nu}\Big)
\nonumber
\\
&+& \frac{1}{2}\,m^2\phi_{0c}\,g_{\mu\nu}
\,-\,\frac{\lambda}{6}\phi_{0c}^3\,g_{\mu\nu}\,.
\label{fra}
\eeq
Finally, we replace here the expansion (\ref{cham}) up to the first
order in curvature, $\phi_{0c}=v_0+v_1$, with $v_0$ and $v_1$ taken
from Eqs. (\ref{saalleaas}) and (\ref{fred}). Also, we use the
normal coordinates expansion of the operator \
$\nabla_{\mu}\nabla_{\nu}-g_{\mu\nu}\Box$ \ which is calculated in
the Appendix A. The final result for the first factor inside the
integral in Eq. (\ref{ind-1}) has the form
\beq
&&
\frac{1}{\sqrt{-g}}\,\frac{\delta S}{\delta g_{\mu\nu}\delta\phi}
\Big|_{\phi_{0c}}
\,=\,
\xi v_0\big(\pa_{\mu}\pa_{\nu}-\eta_{\mu\nu}\pa^2\big)
%% \nonumber \\ &+&
\,+\,\frac{\xi^2 v_0}{2m^2}\,R\,\big(\pa_{\mu}\pa_{\nu} - \eta_{\mu\nu}\pa^2\big)
\,-\,\xi v_0\,R_{\mu\nu}
\label{firstfactor}
\\
&+& \frac{1}{3}\,\xi v_0\,
\Big[2\,R^\la\,_{(\mu\nu)}\,_{\ta}\,y^{\ta}\pa_{\la}
\,+\, 2\,\eta_{\mu\nu}\,\,R^{\la}_{\ta}\,y^{\ta}\pa_{\la}
%% \nonumber \\
\,+\, R_{\mu\al\nu\be}\,y^\al y^\be\,\pa^2
\,+\,\eta_{\mu\nu}
\,R^{\rho}\,_{\al\be}\,^{\sigma}\,y^{\al}y^{\be}\,
\pa_{\rho}\pa_{\si}\Big]\,.
\nonumber
\eeq
The first term in the {\it r.h.s.} is of the zero order in curvature
and the rest of the terms are of the first order in curvature. Indeed,
after all calculations are completed, we will trade the metric
$\eta_{\mu\nu}$ in the point $P$ to the general one $\,g_{\mu\nu}$,
but for a while it is better we write it in the way we did.

Let us now consider the second  factor inside the integral in
Eq. (\ref{ind-1}),
\beq
\Biggl(\frac{1}{\sqrt{-g}}\,
\frac{\de S^2[g,\phi_{0c}]}{\de\phi\,\de\phi}\Biggr)^{-1}_{y,z}
\,=\,G(y,z;\,\phi_{0c})\,,
\label{propag}
\eeq
This is nothing else but the propagator of the scalar excitations
near the point of the minima. It is important to remember that we
will need the dependence on the curvature. Therefore, according
to \cite{BunPar,Parker-Toms} (see also \cite{CorPot}) one has to
modify the (\ref{propag}) to the form
\beq
\Biggl(\frac{1}{[-g(y)]^{1/4}\,[-g(z)]^{1/4} }\,\times\,
\frac{\de S^2[g,\phi_{0c}]}{\de\phi(y)\,\de\phi(z)}\Biggr)^{-1}
\,=\,\,{\bar G}(y,z;\,\phi_{0c})\,.
\label{propag-m}
\eeq
Now we can use the known result for the propagator from the
mentioned references \cite{BunPar,Parker-Toms,CorPot}, but first
we have to evaluate the mass of the scalar excitations near the
point of the minima.

One can start from the full propagator with \ $\phi_{oc}=v_0+v_1$.
Starting from the equation (\ref{esatr}) we arrive at
\beq
\frac{1}{\sqrt{-g}}\frac{\delta^2 S}{\delta\phi\,\delta\phi}=
- \Box + m^2 + \xi R - \frac{\la}{2}\,\phi^2\,.
\label{gums}
\eeq
Next we replace
\beq
\phi \rightarrow \phi_{0c}^2
\,=\,(v_0+v_1)^2 \approx v_0^2+2v_0v_1\,.
\label{sic}
\eeq
Replacing (\ref{fred}) and (\ref{saalleaas}) into (\ref{sic}),
after some small algebra we arrive at
\beq
\frac{1}{\sqrt{-g}}\,\frac{\delta^2 S}{\delta\phi\,\delta\phi}
&=&
- \,\Box + 2m^2
\,+\,\xi\Big(1-\frac{6m^2}{\Box+2m^2}\Big)\,R
\nonumber
\\
&\approx &
- (\Box+2m^2) \,-\,2\xi\,R \,,
\label{trash}
\eeq
where at the last step we used the ${\cal O}(R)$-approximation, as it
was already discussed above. Now, after we compare the last expression
with Eq. (\ref{esatr}), it is clear that (\ref{trash}) means we have
a propagator of a scalar particle with {\it positive} mass \ $2m^2$ \
and with the non-minimal parameter \ $-2\xi$. By using the general
expression (\ref{propa}) we obtain the   Euclidean version of the
second factor inside the integral in Eq. (\ref{ind-1}) in the form
\beq
{\bar G}(z-y)
&=&
\int\frac{d^4k}{(2\pi)^4}\,e^{ik(z-y)}\,
\left[\frac{1}{k^2 + 2m^2}
- \Big(2\xi - \frac16\Big)\,\frac{R}{(k^2 + 2m^2)^2}\right]\,.
\label{propa-2}
\eeq

The third factor of the integrand in Eq. (\ref{ind-1}) is nothing
else but the effective equation of motion (\ref{eps}). According
to (\ref{eps-12}) we can write it as a sum of classical and quantum
parts, $\vp = {\bar \vp}^{(0)} + \hbar{\bar \vp}^{(1)}$, where the
last can be also expanded into series in scalar curvature,
$\,{\bar \vp}^{(1)} = {\bar \vp}^{(1)}_0 \,+\,{\bar \vp}^{(1)}_1$.
For the sake of completeness we have calculated these expressions,
but since they are rather cumbersome, we postpone them to Appendix B.

Now we are in a position to derive
$\langle {\bar T}_{\mu\nu}\rangle_{i}$. As a first step we obtain
the flat-space expression and then consider a bit more complicated
curvature-dependent terms.

In the flat-space limit we have only first terms in the {\it r.h.s.}
of Eq. (\ref{firstfactor}) and (\ref{propa-2}), and also need only
$\,{\bar \vp}^{(1)}_0\,$-parts in the equation of motion (both
divergent and renormalized versions). In this way we arrive at the
expression
\beq
\langle T_{\mu\nu}(x)\rangle_i^0
&=&
2\hbar\,\xi v_0\,
\int d^4z \, d^4y\,\,\de^4(x-y)
\nonumber
\\
&\times&
\big(\pa_\mu\pa_\nu - \eta_{\mu\nu}\pa^2\big)_{y}
\int\frac{d^4k}{(2\pi)^4}
\,\frac{e^{ik(y-z)}}{k^2+2m^2} \,\bar{\vp}^{(1)}_0(z),\,
\label{limaaas}
\eeq
where the upper index $0$ indicates flat-space limit.
After performing integration over $y$ and using
$\,\,\bar{\vp}^{(1)}_0(z)=const$, we get
\beq
\langle T_{\mu\nu}(x)\rangle_i^0
&=&
2\hbar\xi v_0\int d^4z
\,\big(\pa_\mu\pa_\nu - \eta_{\mu\nu}\pa^2\big)_{x}
\int\frac{d^4k}{(2\pi)^4}\frac{e^{ik(x-z)}}{k^2+2m^2}
\,\bar{\vp}^{(1)}_0(z)
\nonumber
\\
&=&
2\hbar\xi \,v_0\,\bar{\vp}^{(1)}_0\,
\big(\pa_\mu\pa_\nu - \eta_{\mu\nu}\pa^2\big)_{x}
\int d^4k\,\,
\frac{e^{ikx}}{k^2+2m^2}
\,\int \frac{d^4z}{(2\pi)^4}\,e^{-ikz}
\nonumber
\\
&=&
2\hbar\xi\, v_0\,\bar{\vp}^{(1)}_0
\int
d^4k\,\,\de^4(k)\,\,
\frac{k_{\mu}k_{\nu}-k^2\eta_{\mu\nu}}{k^2+2m^2}\,\,e^{ikx}
\,=\,0\,.
\label{noca}
\eeq
Thus, the contribution of the last term in (\ref{lavo}) to the
induced cosmological constant is zero.

As a first byproduct we also obtain that in curved space-time
the contributions of the third factor, being it
$\,\bar{\vp}^{(1)}_{1,div}\,$ or
$\,\bar{\vp}^{(1)}_{1,ren}$, are also vanishing. The reason is
that both are constants in the ${\cal{O}}(R)$ approximation and
we did not use an explicit form of a constant $\,\bar{\vp}^{(1)}_0\,$
in the calculation presented above.

As a second byproduct we can see that in curved space-time
the curvature-dependent contribution of the second factor
(\ref{propa-2}) vanish too. The reason is that, if we trade
\beq
\frac{1}{k^2 + 2m^2} \,\,\rightarrow\,\,
- \,\Big(2\xi - \frac16\Big)\,\frac{R}{(k^2 + 2m^2)^2}
\label{propa-trade}
\eeq
in (\ref{limaaas}), the zero output of the integral will obviously
remain the same. So, after all we need to take into account only the
curvature-dependent terms in the first factor, Eq. (\ref{firstfactor}).
\vskip 2mm

The last step is to perform the curved - space calculation in
the $\,{\cal{O}}(R)$ order. Taking into account the arguments
presented above, we arrive at
\beq
\langle T_{\mu\nu}(x)\rangle_i^1
&=&
2\hbar\,\bar{\vp}^{(1)}_{0}\,
\int d^4y d^4z\int\frac{d^4k}{(2\pi)^4}
\sum_{i=1}^5 O^{(i)}_{\mu\nu}(y)\de^4(x-y)
\,\frac{e^{ik(y-z)}}{k^2+2m^2}\,,
\label{i}
\eeq
where
\beq
O^{(1)}_{\mu\nu}
&=& -\,\xi v_0\,R_{\mu\nu}\,,
\nonumber
\\
O^{(2)}_{\mu\nu}
&=&
\frac{\xi^2v_0}{2m^2}\,R\,(\pa_{\mu}\pa_{\nu}-\eta_{\mu\nu}\pa^2)\,,
\nonumber
\\
O^{(3)}_{\mu\nu}
&=&
\frac{2}{3}\,\xi v_0\,\big[R^{\la}\,_{(\mu\nu)\ta}
+\eta_{\mu\nu}R^{\la}_{\ta}\big]\,y^{\ta}\pa_{\la}\,,
\nonumber
\\
O^{(4)}_{\mu\nu}
&=&
\frac{1}{3}\xi v_0\,R_{\mu\al\nu\be}\,y^{\al}y^{\be}\,\pa^2\,,
\nonumber
\\
O^{(5)}_{\mu\nu}
&=&
\frac{1}{3}\,\xi v_0\,\eta_{\mu\nu}\,R_{\al\,\,\,\,\be}^{\,\,\,\rho\si}
\,\,y^{\al}y^{\be}\pa_{\rho}\pa_{\si}\,.
\label{Os}
\eeq
Let us evaluate all the terms of (\ref{i}), indicating the term in
(\ref{Os}) by the left upper index.

The contribution of $O^{(1)}_{\mu\nu}$ has the form which strongly
resembles (\ref{limaaas}) and can be treated in the same way,
\beq
^{(1)}\langle T_{\mu\nu}\rangle^{1}_i
&=& -\,2\hbar\xi\, v_0\,R_{\mu\nu}\,\bar{\vp}^{(1)}_0
\int d^4y\,d^4z\int\frac{d^4k}{(2\pi)^4}\,
\frac{e^{ik(y-z)}}{k^2+2m^2}\,\de^4(x-y)
\nonumber
\\
&=&
-\,2\hbar\xi\, v_0\,R_{\mu\nu}\,\bar{\vp}^{(1)}_{0}
\int d^4k \,\de^4(k)\,\frac{e^{ikx}}{k^2+2m^2}
\,=\,-\,\frac{\hbar\xi v_0}{m^2}\,R_{\mu\nu}\,\bar{\vp}^{(1)}_{0}\,.
\label{O1fin}
\eeq

Next, the contribution of $O_{\mu\nu}^{(2)}$ vanish, for it
has the same structure as the flat-space term (\ref{limaaas}),
\beq
^{(2)}\langle T_{\mu\nu}\rangle^{1}_i &=& 0\,.
\label{O2fin}
\eeq

The contribution of $(O)_{\mu\nu}^{(3)}$ can be presented in the
form
\beq
^{(3)}\langle T_{\mu\nu}\rangle^{1}_i
&=&
\frac{4\hbar\,\xi v_0}{3}\,
\big[R^{\la}\,_{(\mu\nu)\ta} + \eta_{\mu\nu}R^{\la}_{\ta}\big]\,
\bar{\vp}^{(1)}_{0}\,I^\ta_\la\,,
\label{O3}
\eeq
where
\beq
I^\ta_\la
&=&
\int d^4y\,d^4z\int \frac{d^4k}{(2\pi)^4}\,\,\de^4(x-y)
\,y^{\ta}\frac{\pa}{\pa y^{\la}}\,\frac{e^{ik(y-z)}}{k^2+2m^2}\,.
\label{conj}
\eeq
The last integral can be calculated by elementary means to give
\beq
I^\ta_\la
&=&
\frac{1}{2m^2}\,\de ^{\ta}_{\la}
\label{I2}
\eeq
and hence, after some small algebra, we obtain
\beq
^{(3)}\langle T_{\mu\nu}\rangle^{1}_i
&=&
-\,\frac{2\hbar\,\xi\,v_0}{3m^2}
\,\big( R_{\mu\nu} - R \eta_{\mu\nu}\big)\,\bar{\vp}^{(1)}_0\,.
\label{O3fin}
\eeq

The contributions of $(O)_{\mu\nu}^{(4)}$ and $(O)_{\mu\nu}^{(5)}$
can be expressed in the form
\beq
^{(4,5)}\langle T_{\mu\nu}\rangle^{1}_i
&=&
\frac{2\hbar\,\xi v_0}{3}\,
\bar{\vp}^{(1)}_{0}\,\big[R_{\mu\al\nu\be}\,\eta^{\rho\si}
- R_{\al}\,^{\rho}\,_{\be}\,^{\si}\,\eta_{\mu\nu}\big]
\,J_{\rho\si ,}\,^{\al\be}\,,
\label{O45}
\eeq
where
\beq
J_{\rho\si}\,,^{\al\be}
&=&
\int d^4y\,d^4z \int\frac{d^4k}{(2\pi)^4}\,\de^4(x-y)
\,y^\al y^\be\,\frac{\pa^2}{\pa y^{\rho}\pa y^{\si}}
\,\frac{e^{ik(y-z)}}{k^2+2m^2}\,.
\label{J2}
\eeq
Taking this integral we obtain
\beq
J_{\rho\si}\,,^{\al\be}
&=&
\frac{1}{m^2}\de_{\rho\si}\,,^{\al\be}
\quad \mbox{where}\quad
\de_{\rho\si}\,,^{\al\be}
= \frac12\big(\de_\rho^\al \de_\si^\be
- \de_\rho^\be \de_\si^\al\big)\,.
\label{J2fin}
\eeq
Using this result, after small algebra we arrive at
\beq
^{(4,5)}\langle T_{\mu\nu}\rangle^{1}_i
&=&
\frac{2\hbar\,\xi v_0}{3m^2}\,\bar{\vp}^{(1)}_{0}\,
\Big( R_{\mu\nu} + \frac{1}{2} \,R \eta_{\mu\nu}\Big)\,.
\label{O45fin}
\eeq

The total expression can be obtained by replacing the results
for all contributions (\ref{O1fin}), (\ref{O1fin}), (\ref{O3fin})
and (\ref{O45fin}) into (\ref{Os}). After a some algebra we
come to the result
\beq
\langle T_{\mu\nu}\rangle^{1}_i
&=&
\frac{\hbar\,\xi v_0}{m^2}\,
\bar{\vp}^{(1)}_{0}\Big(-\,R_{\mu\nu}\,+\, R\eta_{\mu\nu}\Big)\,.
\label{i-fin}
\eeq
Obviously, this expression is different from $G_{\mu\nu}$ and
therefore it violates covariance and conservation law. However, if
we sum up with the previous result $\langle T_{\mu\nu}\rangle^{1}_v$
from Eq. (\ref{lavo10}), we arrive at the expression which agrees
with our expectations,
\beq
\langle T_{\mu\nu}\rangle^{1}
&=&
\langle T_{\mu\nu}\rangle^{1}_i
\,+\,\langle {\bar T}_{\mu\nu}\rangle_{v}^1
\nonumber
\\
&=&
-\,2\,\hbar V_1(v_0)\,G_{\mu\nu}
\,+\,\hbar V_0(v_0)\,g_{\mu\nu}
\,-\,\frac{\hbar \,\xi\,v_0}{m^2}\,
\Big(R_{\mu\nu}\,-\,\frac12\, Rg_{\mu\nu}\Big)
{\bar \vp}^{(1)}_0\,,
\nonumber
\\
&=&
-\,\hbar \Big[ 2\,V_1(v_0)
\,+\,\frac{\xi\,v_0}{m^2}\,{\bar \vp}^{(1)}_0\Big]\,
G_{\mu\nu}
\,+\,\hbar V_0(v_0)\,g_{\mu\nu}\,,
\label{Tmn-quantum}
\eeq
where we finally replaced the flat metric $\eta_{\mu\nu}$ by
the general one $\,g_{\mu\nu}$.

In order to rewrite the quantum contribution in the final form,
one needs the expressions for $V_0(v_0)$, $V_1(v_0)$ and
${\bar \vp}^{(1)}_0$. The renormalized and divergent versions of
the first two can be obtained from (\ref{REN-potya}) and
(\ref{potya-div}) in the form
\beq
V_0^{ren}(v_0)
&=& \frac{1}{(4\pi)^2}
\,m^4\,\ln \Big(\frac{2m^2}{\mu^2}\Big)\,,
\label{V0-v0-ren}
\\
V_0^{div}(v_0)
&=& \frac{m^2}{32\,\pi^2}
\,\Big[3\Om^2 - 2m^2\,\ln \Big(\frac{\Om^2}{m^2}\Big)\Big]\,.
\label{V0-v0-div}
\eeq
and
\beq
V_1^{ren}(v_0)
&=& -\,\frac{m^2}{(4\pi)^2}\,\ln \Big(\frac{2m^2}{\mu^2}\Big)\,,
\label{V1-v0-ren}
\\
V_1^{div}(v_0)
&=& \frac{1}{32\,\pi^2}\,\Big(\xi-\frac16\Big)\,
\Big[-\,\Om^2 \,+\, 2m^2\,\ln \Big(\frac{\Om^2}{m^2}\Big)\Big]\,.
\label{V1-v0-div}
\eeq

Taking into account in (\ref{Tmn-quantum}) the expressions
for ${\bar \vp}^{(1)}_0$
derived in Eqs. (\ref{B5}) and (\ref{B6}) of Appendix B, we
arrive at the final result for quantum contributions to EMT,
\beq
\langle T_{\mu\nu}\rangle_{ren}
&=&
\frac{\hbar\,m^4}{(4\pi)^2}\,
\ln \Big(\frac{2m^2}{\mu^2}\Big)\,g_{\mu\nu}
\,-\,
\frac{m^2}{(4\pi)^2}
\Big[2(1+3\xi)\,\ln \Big(\frac{2m^2}{\mu^2}\Big)
\,+\,3\xi \Big]\,G_{\mu\nu}
\label{Tmn-q-f-ren}
\eeq
for the renormalized expression and

\newpage
\beq
\langle T_{\mu\nu}\rangle_{div}
&=&
\frac{\hbar \,m^2}{32\,\pi^2}
\,\Big[3\Om^2 - 2m^2\,a\Big(\frac{\Om^2}{m^2}\Big)\Big]\,g_{\mu\nu}
 \nonumber
 \\
 &+&
\frac{\hbar}{16\,\pi^2}\,\Big(4\xi-\frac16\Big)\,
\Big\{\Om^2 \,-\,  2m^2\,\ln{\frac{\Om^2}{m^2}}\Big\}\,
G_{\mu\nu}
\label{Tmn-q-f-div}
\eeq
for the divergent one.

%%%%%%%%%%%%%%%%%%%%%%%%%%%%%%%%%%%%%%%%%%%%%%%%%%%%%%%%%%%
%%%%%%%%%%%%%%%%%%%%%%%%%%%%%%%%%%%%%%%%%%%%%%%%%%%%%%%%%%%
%%%%%%%%%%%%%%%%%%%%%%%%%%%%%%%%%%%%%%%%%%%%%%%%%%%%%%%%%%%
\section{Conclusions}
%%%%%%%%%%%%%%%%%%%%%%%%%%%%%%%%%%%%%%%%%%%%%%%%%%%%%%%%%%%%%%

We have considered several aspects of the Energy-Momentum Tensor
(EMT) of vacuum in curved space-time. A naive calculation using a
momentum cut-off produces a result which apparently violates
general covariance. It was noticed long ago that this is the
effect of the non-covariant cut-off scheme and therefore can
be hardly regarded to be a physical feature of the theory. The
two questions naturally arise in this respect, namely whether
it is possible to introduce a cut-off on a covariant way and
whether it is possible to have a non-trivial quantum contributions
to the Energy-Momentum Tensor of Vacuum.

We have addressed the first and in part the second issue on the
basis of effective action method. In both cases the output of
our investigation was perfectly consistent with the general
expectations based on the known structure of renormalization
in curved space-time and conservation law for the EMT. The
calculations in the theory with SSB have shown some new term
which was unnoticed until now. At the same time, after performing
explicit calculations of this term we have found that the final
results, Eqs. (\ref{Tmn-q-f-ren}) and (\ref{Tmn-q-f-div}), have
the usual form and that the quantum effects always lead only to
the renormalization of the inverse Newton constant and cosmological
constant in Eqs. (\ref{classic}) and in the relations such as
(\ref{Ein-1}) and (\ref{Ein-2}).

It is important to note that our calculations can not be interpreted
as a no-go theorem for a non-trivial quantum contributions to the
low-energy sector of the gravitational action. As it was previously
explained in \cite{DCCR}, the chance to meet such corrections exists,
but this can be verified only in the framework of some qualitatively
new mathematical tool which should not be based on the perturbative
expansion in curvatures. The linear in curvature approximation which
was adopted here does not provide any information about these type
of corrections. However, it was definitely worthwhile to check that
the standard considerations really work for the non-trivial physical
situations such as gravity combined with SSB.

%%%%%%%%%%%%%%%%%%%%%%%%%%%%%%%%%%%%%%%%%%%%%%%%%%%%%%%%%%%
%%%%%%%%%%%%%%%%%%%%%%%%%%%%%%%%%%%%%%%%%%%%%%%%%%%%%%%%%%%
%%%%%%%%%%%%%%%%%%%%%%%%%%%%%%%%%%%%%%%%%%%%%%%%%%%%%%%%%%%
\acknowledgments
The authors thank M. Maggiore for useful discussions of the papers
\cite{Magg1,Magg2}. M.A. was partially supported by Spanish
DGIID-DGA grant 2009-E24/2 and  MICINN grants FPA2009-09638 and
CPAN-CSD2007-00042. The work of P.L. is supported by  the LRSS
grant 224.2012.2  as well as by the RFBR
grant 12-02-00121 and the RFBR-Ukraine grant 11-02-90445.
Some part of this work was done during the visit
of P.L. to UFJF and we are grateful to FAPEMIG for
supporting this visit. Also, B.R. and I.Sh. are grateful to
CAPES, CNPq and FAPEMIG for partial support of their work.
%%%%%%%%%%%%%%%%%%%%%%%%%%%%%%%%%%%%%%%%%%%%%%%%%%%%%%
%%%%%%%%%%%%%%%%%%%%%%%%%%%%%%%%%%%%%%%%%%%%%%%%%%%%%%%%%%%%%%

\section*{Appendix A. \ \ Local momentum representation}
%%%%%%%%%%%%%%%%%%%%%%%%%%%%%%%%%%%%%%%%%%%%%%%%%%%%%%%%%%%%%%
%% \subsection{Review of the Mathematical tools: normal coordinates}

The calculations in Sect. 6 were done using Riemann normal
coordinates (see, e.g., \cite{Petrov} for introduction) and
the local momentum representation technique (see, e.g.,
\cite{BunPar,Parker-Toms}).
In this Appendix we present some necessary elements of these tools
and also derive the operator $g_{\mu\nu}\Box-\na_{\mu}\na_{\nu}$
because it is related to relatively trivial calculations.

The normal coordinates expansion performs around one special
point $P$ (that is related methods work well only for deriving
local quantities), where metric is supposed to be flat Minkowski
one. However, the derivatives of the metric, starting from the
second one, are of course non-zero. The expression for the
metric is
\beq
g_{\mu\nu}=\eta_{\mu\nu}
-\frac{1}{3}R_{\mu\al\nu\be}\,y^{\al}y^{\be}
+ \,...\,.
\label{petros}
\eeq
Here and below all components of the curvature tensor correspond
to the point $P$, also in (\ref{petros}) we have omitted the
higher order terms in curvature tensors and their derivatives.
Furthermore $y^\al$ represent deviation from the point $P$,
such that all partial derivatives below are taken with respect
to $y^\al$. It is fairly easy to derive, using (\ref{petros}),
the following expansions:
\beq
g^{\mu\nu} &=& \eta^{\mu\nu}
+ \frac{1}{3}\,R^{\mu}\,_{\al}\,^{\nu}\,_{\be}\,y^{\al}y^{\be}\,,
\nonumber
\\
\Gamma^{\la}_{\mu\nu} &=& -\,\frac{2}{3}\,R^{\la}
\,_{(\mu\nu)}\,_{\ta}y^{\ta}\,.
\label{Ga}
\eeq
Then for the two covariant derivatives acting on scalar we obtain
\beq
\na_{\mu}\na_{\nu}\,=\,\pa_{\mu}\pa_{\nu}
+\frac{2}{3}\,R^{\la}\,_{(\mu\nu)}\,_{\ta}\,\,y^{\ta}\pa_{\la}\,.
\eeq
Making contraction with
\beq
g^{\mu\nu}=\eta^{\mu\nu}
+\frac{1}{3}\,R^{\mu}\,_{\la}\,^{\nu}\,_{\be}\,\,y^{\al}y^{\be}\,,
\label{2der}
\eeq
we get
\beq
\Box=g^{\mu\nu}\na_{\mu}\na_{\nu}
\,=\,\pa^2
-\frac{2}{3}\,R^{\la}_{\ta}\,\,y^{\ta}\pa_{\la}
+\frac{1}{3}\,R^{\mu}\,_{\al}\,^{\nu}\,_{\be}
\,\,y^{\al}y^{\be}\pa_{\mu}\pa_{\nu}\,,
\label{box}
\eeq
where $\pa^2=\eta^{\mu\nu}\pa_{\mu}\pa_{\nu}$.
Finally, the first operator of our interest is
\beq
\na_{\mu}\na_{\nu}-g_{\mu\nu}\Box
&=&
\pa_{\mu}\pa_{\nu}+\frac{2}{3}\dot{R}^{\la}
\,_{(\mu\nu)}\,_{\ta}y^{\ta}\pa_{\la}-\eta_{\mu\nu}\pa^2
+\frac{2}{3}\eta_{\mu\nu}\dot{R}^{\la}_{\ta}y^{\ta}\pa_{\la}
\nonumber
\\
&-&\frac{1}{3}\eta_{\mu\nu}\dot{R}^{\rho}\,_{\al}\,^{\sigma}
\,_{\be}\,y^{\al}y^{\be}\pa_{\rho}\pa_{\sigma}
-\frac{1}{3}\dot{R}_{\mu\al\nu\be}y^{\al}y^{\be}\pa^2\,.
\label{opa}
\eeq

The next operator we are interested in is the propagator
of scalar field. Direct calculations using (\ref{box})
(see, e.g., \cite{BunPar,Parker-Toms}) lead to the
relevant expression for the propagator of the field of
the mass $m$ in the linear in curvature approximation,
\beq
\bar{G}(y)=\int\frac{d^4k}{(2\pi)^4}e^{iky}
\Big[\frac{1}{k^2+m^2}
- \Big(\xi - \frac16\Big) \, \frac{R}{(k^2+m^2)^2}\Big]\,,
\label{propa}
\eeq
where we already assumed Wick rotation to Euclidean space.
The Eq. (\ref{propa}) was recently used in \cite{CorPot} to
derive the effective potential of scalar field in the
momentum cut-off regularization. Due to the use of
local momentum representation (\ref{propa}) the result has
covariant form, despite the naive application of the cut-off
scheme is supposed to break down even Lorentz invariance.

%%%%%%%%%%%%%%%%%%%%%%%%%%%%%%%%%%%%%%%%%%%%%%%%%%%%%%%%%%%%%%
\section*{Appendix B. \ \ Effective equations of motion}
%%%%%%%%%%%%%%%%%%%%%%%%%%%%%%%%%%%%%%%%%%%%%%%%%%%%%%%%%%%%%%
%% \subsection{Review of the Mathematical tools: normal coordinates}

Here we present the effective equations of motion on the
${\cal O}(R)$-approximation, when the effective action is
reduced to (\ref{GV}).
By using Eqs. (\ref{GV}) and (\ref{eps-12}) we obtain
\beq
\bar{\vp}^{(1)}_{0} &=& -\,\frac{\pa\bar{V}_0}{\pa \,\phi}\Big|_{v_0}
\label{B1}
\eeq
and
\beq
\bar{\vp}^{(1)}_{1}=
-\frac{\pa^2\bar{V}_0}{\pa \,\phi^2}\Big|_{v_0}\cdot v_1
-R\frac{\pa\bar{V}_1}{\pa\phi}\Big|_{v_0}\,.
\label{B2}
\eeq

Let us denote the curvature-independent and mass-independent
part of classical potential as
\beq
V = V(\phi) = \frac{\la}{4\!}\,\phi^4\,.
\label{clapot}
\eeq
From the quantities $\bar{V}_{0}$ and $\bar{V}_{1}$ given by
(\ref{potya-div}) and (\ref{scalar div}) one can easily get
\beq
\frac{\pa\bar{V}^{div}_{0}}{\pa\phi}
&=&
\frac{1}{32\pi^2}\Big[\Om^2V'''-(V''-m^2)V'''
\ln{\frac{\Om^2}{m^2}}\Big]\,,
\nonumber
\\
\frac{\pa^2\bar{V}^{div}_{0}}{\pa\phi^2}
&=& \frac{1}{32\pi^2}\Big[\Om^2V''''-(V''')^2\ln{\frac{\Om^2}{m^2}}
-V''''(V''-m^2)\ln{\frac{\Om^2}{m^2}}\Big]\,,
\nonumber
\\
\frac{\pa\bar{V}^{div}_{1}}{\pa\phi}
&=&
\frac{1}{32\pi^2}\Big(\xi-\frac{1}{6}\Big)
V'''\ln{\frac{\Om^2}{m^2}}\,.
\label{B3}
\eeq

Furthermore, from $\bar{V}^{(0)}_{ren}$ and $\bar{V}^{(1)}_{ren}$
in (\ref{REN-potya}) we obtain
\beq
\frac{\pa\bar{V}^{ren}_{0}}{\pa\,\phi}
&=&
\frac{1}{32\pi^2}(V''-m^2)V'''\Big[
\ln{\Big(\frac{V''-m^2}{\mu^2}\Big)}+\frac{1}{2}\Big]\,,
\nonumber
\\
\frac{\pa^2\bar{V}^{ren}_{0}}{\pa\,\phi^2}
&=&
\frac{1}{32\pi^2}\Big\{\Big[(V''')^2+(V''-m^2)V''''\Big]
\Big[
\ln{\Big(\frac{V''-m^2}{\mu^2}}\Big)+\frac{1}{2}\Big]
+(V''')^2\Big\}\,,
\nonumber
\\
\frac{\pa\bar{V}^{ren}_{1}}{\pa\,\phi}
&=&
-\,\frac{1}{32\pi^2}\Big(\xi-\frac{1}{6}\Big)V'''
\Big[\ln{\Big(\frac{V''-m^2}{\mu^2}\Big)}+1\Big]\,.
\label{B4}
\eeq

Next, we calculate the on-shell expressions by replacing
\ $\phi \to \phi_{0c}$ \ and $\,\la v_0^2=6m^2$, in the form
\beq
\frac{\pa\bar{V}^{div}_{0}}{\pa\phi}\Big|_{v_0}
&=&\frac{1}{32\pi^2}\Big[\la v_0\Om^2
-2\la m^2v_0\ln{\frac{\Om^2}{m^2}}\Big]\,,
\nonumber
\\
\frac{\pa\bar{V}^{div}_{1}}{\pa\phi}\Big|_{v_0}
&=&\frac{1}{32\pi^2}\Big(\xi-\frac{1}{6}\Big)
\la v_0\ln{\frac{\Om^2}{m^2}}\,,
\nonumber
\\
\frac{\pa^2\bar{V}^{div}_{0}}{\pa\phi^2}\Big|_{v_0}
&=& \frac{1}{32\pi^2}\Big[\la\,\Om^2
-8\la m^2\ln{\frac{\Om^2}{m^2}}\Big]\,.
\label{ders}
\eeq

Similarly, the analogous renormalized on-shell expressions are
\beq
\frac{\pa\bar{V}^{ren}_{0}}{\pa\phi}\Big|_{v_0}
&=& \frac{1}{16\pi^2}\la\, m^2 v_0\Big[
\ln{\Big(\frac{2m^2}{\mu^2}}\Big)+\frac{1}{2}\Big]\,,
\nonumber
\\
\frac{\pa^2\bar{V}^{ren}_{0}}{\pa\phi^2}\Big|_{v_0}
&=&\frac{\la\,m^2}{16\pi^2}\Big[4
\ln{\Big(\frac{2m^2}{\mu^2}}\Big)+5\Big]\,,
\nonumber
\\
\frac{\pa\bar{V}^{ren}_{1}}{\pa\phi}\Big|_{v_0}
&=&-\frac{\la v_0}{32\pi^2}\Big(\xi-\frac{1}{6}\Big)
\Big[\ln{\Big(\frac{2m^2}{\mu^2}\Big)}+1\Big]\,.
\label{ders-ren}
\eeq

At this point we can derive the elements of equations of motion,
\beq
\bar{\vp}^{(1)}_{0,div}
&=&
\frac{\pa\bar{V}^{(0)}_{div}}{\pa\phi}\Big|_{v_0}
=-\frac{\la v_0}{32\pi^2}\Om^2
+\frac{\la\,m^2 v_0}{16\pi^2}
\ln{\frac{\Om^2}{m^2}}\,,
\label{B5}
\\
\bar{\vp}^{(1)}_{0,ren}
&=&
\frac{\pa\bar{V}^{(0)}_{ren}}{\pa\phi}\Big|_{v_0}
=-\frac{1}{16\pi^2}\la\,m^2v_0
\Big[\ln{\Big(\frac{2m^2}{\mu^2}\Big)}+\frac{1}{2}\Big]\,,
\label{B6}
\\
\bar{\vp}^{(1)}_{1,div}
&=&
-\frac{\pa^2\bar{V}_0^{div}}{\pa \,\phi^2}\Big|_{v_0}\cdot v_1
-R\,\frac{\pa\bar{V}_1^{div}}{\pa\phi}\Big|_{v_0}
\nonumber
\\
&=&
-\frac{3\xi}{32\pi^2v_0}\Om^2 R
+\frac{\la v_0}{32\pi^2} \Big(3\xi
+\frac{1}{6}\Big)\,R\,\ln{\frac{\Om^2}{m^2}}\,,
\label{B7}
\\
\bar{\vp}^{(1)}_{1,ren}
&=&
-\frac{\pa^2\bar{V}_0^{ren}}{\pa \,\phi^2}\Big|_{v_0}\cdot v_1
\,-\,R\,\frac{\pa\bar{V}_1^{ren}}{\pa\phi}\Big|_{v_0}
\nonumber
\\
&=&
-\frac{\la v_0}{32\pi^2}
\Big(3\xi+\frac{1}{6}\Big)\,R\,\ln{\Big(\frac{2m^2}{\mu^2}\Big)}
- \Big(4\xi+\frac{1}{6}\Big)\frac{\la v_0}{32\pi^2}\,R\,.
\label{B8}
\eeq
The last observation is that, as we have already mentioned in the
main text, all these expressions must be treated as constants in
the given approximation.
%%%%%%%%%%%%%%%%%%%%%%%%%%%%%%%%%%%%%%%%%%%%%%%%%%%%%%
\vskip 10mm
%%%%%%%%%%%%%%%%%%%%%%%%%%%%%%%%%%%%%%%%%%%%%%%%%%%%%%
\renewcommand{\baselinestretch}{0.9}

\end{document}